\documentclass[twocolumn,apj]{emulateapj}
\usepackage{amsmath,amssymb}
\usepackage{graphicx}
\usepackage{booktabs}
\usepackage{bm}
\usepackage{url}
\providecommand{\nolinkurl}[1]{\url{#1}}

\newcommand{\Rsun}{R_\odot}
\newcommand{\kms}{\mathrm{km\,s^{-1}}}

\shorttitle{MHD Modulation of Streamer Plasma Release}
\shortauthors{Podladchikova}
\submitted{Accepted for publication in The Astrophysical Journal, 4 September 2026}

\begin{document}

\title{Can MHD Oscillations Modulate Quasi-Periodic Plasma Release from Coronal Streamers?}

\author{O. Podladchikova}
\affil{Igor Sikorsky Kyiv Polytechnic Institute, Kyiv, Ukraine}
\affil{Leibniz Institute for Astrophysics Potsdam, Germany}
\email{epodlad@gmail.com}

\begin{abstract}
Periodic density structures in the slow solar wind are commonly associated with quasi-periodic plasma release from coronal streamer and open--closed boundary regions, but the origin of their temporal organization remains uncertain. We distinguish the process that releases plasma from the process that sets or modifies its cadence. In the proposed source--modulator framework, S-Web, interchange, cusp, or current-sheet reconnection releases the plasma, while structured MHD responses may modulate density or release rate. An observation-informed parameter-space analysis gives compact slow-mode periods of about $22$--$231$ min; reproducing $80$--$130$ min requires an effective path length of $0.35$--$1.20\Rsun$. In a classical top-hat cylinder with $L/a=5$--$50$ and $\rho_i/\rho_e=1.6$--$2.5$, the fundamental fast sausage mode is not trapped and the minimum trapped harmonic is approximately $n=4$--$50$. A transverse fast-timescale benchmark based on $a=0.10$--$0.45\Rsun$ and outer-coronal fast speeds of $250$--$530\ \kms$ gives $5.7$--$55$ min. A corresponding fast-interface proxy gives $4.4$--$93$ min, but is not a current-sheet eigenmode solution. Kink responses are constrained empirically rather than by inserting observer-frame propagation speeds into a standing-mode formula: one COR1 event showed an approximately $25$ min pulse, whereas a LASCO/COR2 survey found global streamer-wave periods of $2$--$8$ hr, observer-frame speeds of $360$--$740\ \kms$, and typically only one or two visible cycles. Among the candidates considered here, compact slow modes are therefore the best-constrained stable compressive modulators; confirming their role requires further constraints on longitudinal reflection and damping. Surface/current-sheet responses remain high-priority coupling candidates but require a sheet-specific dispersion relation; kink responses are transient geometric candidates; and tearing/plasmoid formation remains a strong intrinsic reconnection-driven alternative.
\end{abstract}


\section{Introduction}

Periodic density structures (PDS) are quasi-periodic trains of enhanced density that are advected with the solar wind. They are not, in general, propagating density waves. A spacecraft samples them as a time series because the solar-wind flow carries successive structures across the instrument. Remote white-light observations can reveal the larger members of this population as streamer blobs or repeated density enhancements, whereas smaller structures may be detected only in situ \citep{Viall2009,Viall2010,StansbyHorbury2018,Kepko2024}.

The extension of white-light observations from coronagraphic to heliospheric fields of view allows recurrent density structures to be followed over a much larger radial range \citep{Viall2010,ViallVourlidas2015}. This observational continuity motivates a source-region question: can an MHD response established in the structured corona leave a measurable imprint on the cadence of advected PDS, even though the PDS themselves are not waves? The present work addresses this question without assuming that every observed recurrence requires an oscillatory clock.

The observed PDS frequency range is broad. A statistical analysis of Wind data found significant structures over approximately $0.1$--$5$ mHz, or about $3.3$--$167$ min. It also found two different occurrence distributions: a high-frequency enhancement between $1$ and $3$ mHz, centered near $2.1$ mHz ($\simeq7.9$ min), and a lower-frequency enhancement below $1$ mHz, centered near $0.2$ mHz ($\simeq83$ min) \citep{Kepko2024}. Individual remote-sensing studies have also reported periods from roughly $100$ min to several hours \citep{Viall2010,ViallVourlidas2015}. These results do not support one universal period. They instead suggest that several source regions or physical mechanisms may contribute.

Reconnection-based models provide a natural source of structured slow-wind plasma. In the S-Web model, narrow open-field corridors map into a broad web of separatrices and quasi-separatrix layers. Interchange reconnection along open--closed boundaries can release plasma with closed-corona composition onto open magnetic field lines \citep{Antiochos2011}. Reconnection and tearing near streamer cusps and the heliospheric current sheet can also release blobs and flux ropes \citep{Chen2009,Reville2020}. These models explain how plasma is released. They do not, by themselves, uniquely determine why the release should occur with particular quasi-periodic timescales.

This paper investigates a related question: can structured MHD responses of streamer, cusp, separatrix, or current-sheet systems organize the release in time? The proposed oscillations are not required to carry a density enhancement across closed magnetic field lines. Instead, a mode may directly modulate density and pressure, or it may change the cusp geometry, current-sheet thickness, inflow speed, and reconnection rate. Reconnection then produces the released plasma structures.

Reconnection itself can also be quasi-periodic. Tearing, plasmoid formation, and nonlinear current-sheet dynamics can generate repeated release without an external clock. We therefore do not assume that an MHD modulator is required. The reason to test one is narrower: reconnection-based models explain the release channel, but may not uniquely specify the observed cadence in every configuration. A structured MHD response could introduce frequency selection or could modulate the timing, amplitude, or spatial localization of an already established reconnection process.

The study is analytical. Its purpose is not to provide a unique identification of the MHD mode responsible for PDS. Instead, we determine whether physically plausible coronal structures can support oscillations with timescales and lifetimes compatible with the observations. We use analytical estimates to constrain geometry, phase speed, density contrast, background flow, trapping, leakage, and damping. This provides a physically defined parameter space for future observation-constrained MHD simulations and forward modeling.

The purpose of this work is not to develop a seismological model of coronal streamers. Established MHD waveguide and seismological concepts are used only as analytical tools to test whether candidate responses can survive long enough and couple strongly enough to affect plasma release. The central question is therefore not whether MHD oscillations replace reconnection, but whether they can modulate the timing, amplitude, or spatial localization of reconnection-driven plasma release.

The novelty of the present study lies in separating the origin of plasma release from the origin of its cadence. Instead of asking whether one MHD eigenmode generates all PDS, we place intrinsic tearing/plasmoid cycles and MHD-modulated reconnection in the same analytical framework. Each candidate is evaluated using the same physical constraints: period, realistic geometry and plasma parameters, trapping or leakage, damping or growth time, and coupling to density or reconnection. The candidates are then compared on physical grounds rather than selected by period matching alone. This approach identifies where MHD modulation is physically allowed, where intrinsic reconnection may be sufficient, and where the two pathways may coexist.

The framework also leads to distinct, testable predictions. A compact slow-mode contribution should produce correlated density, temperature or pressure, and field-aligned velocity perturbations near the cusp, with periods that scale with effective path length and tube speed. A sausage contribution should produce coherent width--density--white-light brightness variations and may show dispersive or rapidly damped behavior. Kink modulation should show transverse displacement with weak intrinsic compression, whereas surface or current-sheet modulation should correlate changes in sheet position or thickness, cusp geometry, or inflow with release timing. In contrast, an intrinsic tearing/plasmoid cycle should produce plasmoid or flux-rope sequences without requiring a coherent upstream oscillatory signature. Coordinated coronagraphic, EUV or spectroscopic, and in situ observations can therefore reject incompatible mechanisms even when their periods overlap.

\begin{figure*}
\centering
\includegraphics[width=0.98\textwidth]{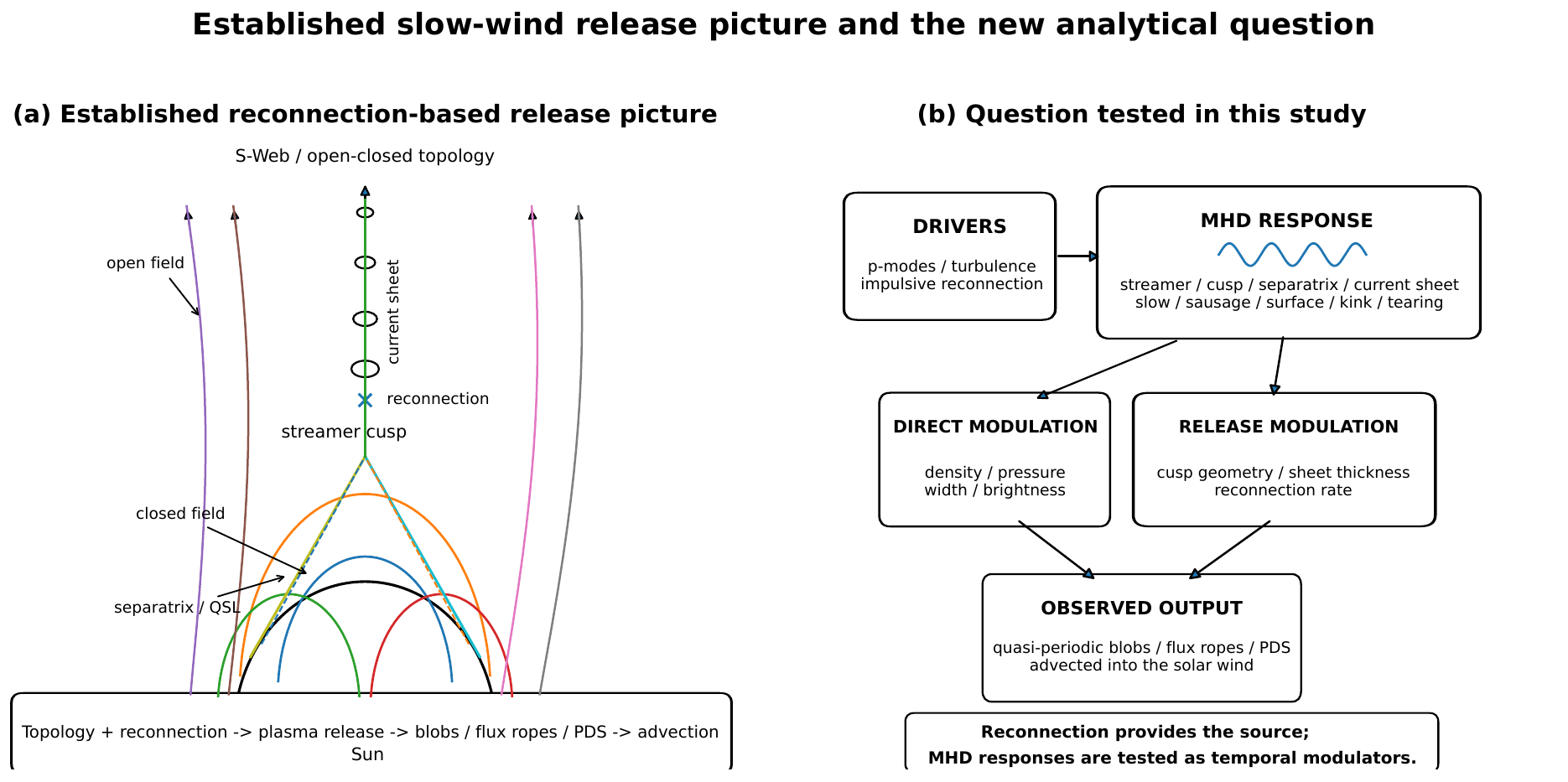}
\caption{Established slow-wind release picture and the question tested here. (a) In the reconnection-based picture, S-Web topology, separatrices, quasi-separatrix layers, and streamer cusp/current-sheet reconnection release plasma from closed or open--closed boundary regions. The resulting blobs, flux ropes, and density enhancements are advected outward and observed as PDS. (b) The present study asks whether structured MHD responses can directly modulate density or can modulate cusp and current-sheet reconnection, thereby organizing the release in time. Reconnection remains the plasma source; MHD responses are tested as possible temporal modulators.}
\label{fig:overview}
\end{figure*}

\section{Observational Constraints on Periodic Density Structures}

\subsection{Remote-sensing and in situ manifestations}

White-light coronagraphs and heliospheric imagers detect electron-density enhancements through Thomson-scattered light. Large streamer blobs and trains of density enhancements can therefore be followed outward from the corona. STEREO/SECCHI observations showed periodic density structures entering and moving through the heliospheric-imager field of view, supporting a coronal origin for at least part of the in situ PDS population \citep{Viall2010}. Composition measurements provide an independent constraint because abundance variations are established in the low corona and remain frozen into the expanding wind \citep{Viall2009,Kepko2024}.

PDS should not be identified only with bright classical streamer blobs. In situ measurements also reveal smaller density enhancements, pressure-balanced structures, and structures with possible flux-rope signatures. Helios measurements inside $0.6$ AU showed quasiperiodic density structures with accompanying temperature changes, consistent with formation during solar-wind release \citep{DiMatteo2019}. Other Helios observations found density structures over a wide range of radial scales, including structures too small for current white-light imagers to resolve \citep{StansbyHorbury2018}.

\subsection{Observed period populations}

For the present analysis, we use three broad observational constraints rather than a fixed sequence of periods:
\begin{enumerate}
\item a high-frequency PDS population near $5.6$--$16.7$ min, corresponding to the $1$--$3$ mHz enhancement;
\item a low-frequency population from tens of minutes to about $167$ min, with a broad maximum near $80$--$90$ min;
\item longer remote or model-produced structures with periods of several hours.
\end{enumerate}
The boundaries are not sharp. The first two intervals are statistical populations, not discrete eigenfrequencies, and individual events can occur outside them \citep{Kepko2020,Kepko2024}.

\subsection{What period matching can and cannot establish}

Agreement between an observed period and a travel-time estimate is necessary but not sufficient for mode identification. A candidate must also satisfy four conditions: (1) the required geometry and plasma parameters must be realistic; (2) the mode must be trapped, partially trapped, or long-lived enough to produce several release cycles; (3) it must couple to density or reconnection; and (4) its predicted phase relations or morphological signatures must be observable. The same period can be produced by different combinations of length, phase speed, harmonic number, and flow. Period matching is therefore used here as a parameter constraint, not as proof of a mechanism.

\section{Source--Modulator--Output Framework}

Figure~\ref{fig:overview} separates the established plasma-release picture from the new modulation question and distinguishes three physical roles. The \emph{source} determines where plasma is released. The \emph{modulator} determines whether density or release rate develops a preferred or quasi-periodic timescale. The \emph{output} is a train of plasma structures that is carried outward by the background flow.

This separation removes a difficulty in a direct resonator interpretation. A standing wave in a closed structure does not need to transport a blob across closed field lines. Instead, pressure perturbations in a compact arcade may reach the cusp, or transverse and surface motions may change the local reconnection conditions. A simple representation is
\begin{equation}
R_{\rm rec}(t)=R_0\left[1+\epsilon\cos(\omega t+\phi)\right],
\end{equation}
where $R_{\rm rec}$ is a measure of the reconnection or release rate. This equation is only a phenomenological description. The coupling coefficient $\epsilon$ must ultimately be obtained from MHD modeling or observations.

\section{Candidate Magnetic Structures and Boundary Conditions}

\subsection{Compact cusp cavity and closed arcade}

A compact closed arcade below a streamer cusp can provide two longitudinal reflection regions and may support standing slow magnetoacoustic oscillations. The word \emph{compact} distinguishes this geometry from a slow mode extending along the full length of a giant streamer. The effective length is the field-aligned distance between the relevant reflection or coupling regions, not automatically the full visible streamer length.

The streamer cusp is physically important as a transition and release region, but it should not be treated as a universal rigid cavity with one characteristic length and one fundamental eigenperiod. Its height, geometry, magnetic connectivity, surrounding flow, and current-sheet properties depend on the global and local coronal configuration. Candidate oscillatory timescales must therefore be related to observationally constrained local or sub-streamer structures rather than assigned to an idealized cusp of fixed size.

\subsection{Streamer body and dense plasma sheet}

A dense streamer stalk or plasma sheet can act as a fast-mode waveguide. Its finite transverse width, density contrast, and external Alfv\'en speed control whether sausage- or kink-like disturbances are trapped, dispersive, or leaky. Coronal streamers have already been treated as structured MHD waveguides for seismological applications. Observed kink motions have been used to estimate the magnetic-field profile in the extended streamer corona \citep{Srivastava2016}.

\subsection{Separatrix and quasi-separatrix boundaries}

Separatrices and quasi-separatrix layers divide magnetic domains with different connectivity. Surface-like disturbances can be concentrated near these boundaries. Such modes are relevant because the boundaries are also natural locations for current formation and interchange reconnection. A boundary mode may therefore have weak volume compression but strong coupling to the release process.

\subsection{Current sheet above the streamer cusp}

The current sheet above the cusp contains strong transverse gradients of magnetic field, density, and flow. It can support guided or surface-like waves, kink- or sausage-like deformations, and unstable tearing or shear-driven modes. These processes can change current-sheet thickness, reconnection inflow, and plasmoid formation. Streamer observations have also shown kink-like disturbances associated with shear-flow instability \citep{Feng2013}.

\section{Candidate MHD Responses}

\begin{figure*}
\centering
\includegraphics[width=0.96\textwidth]{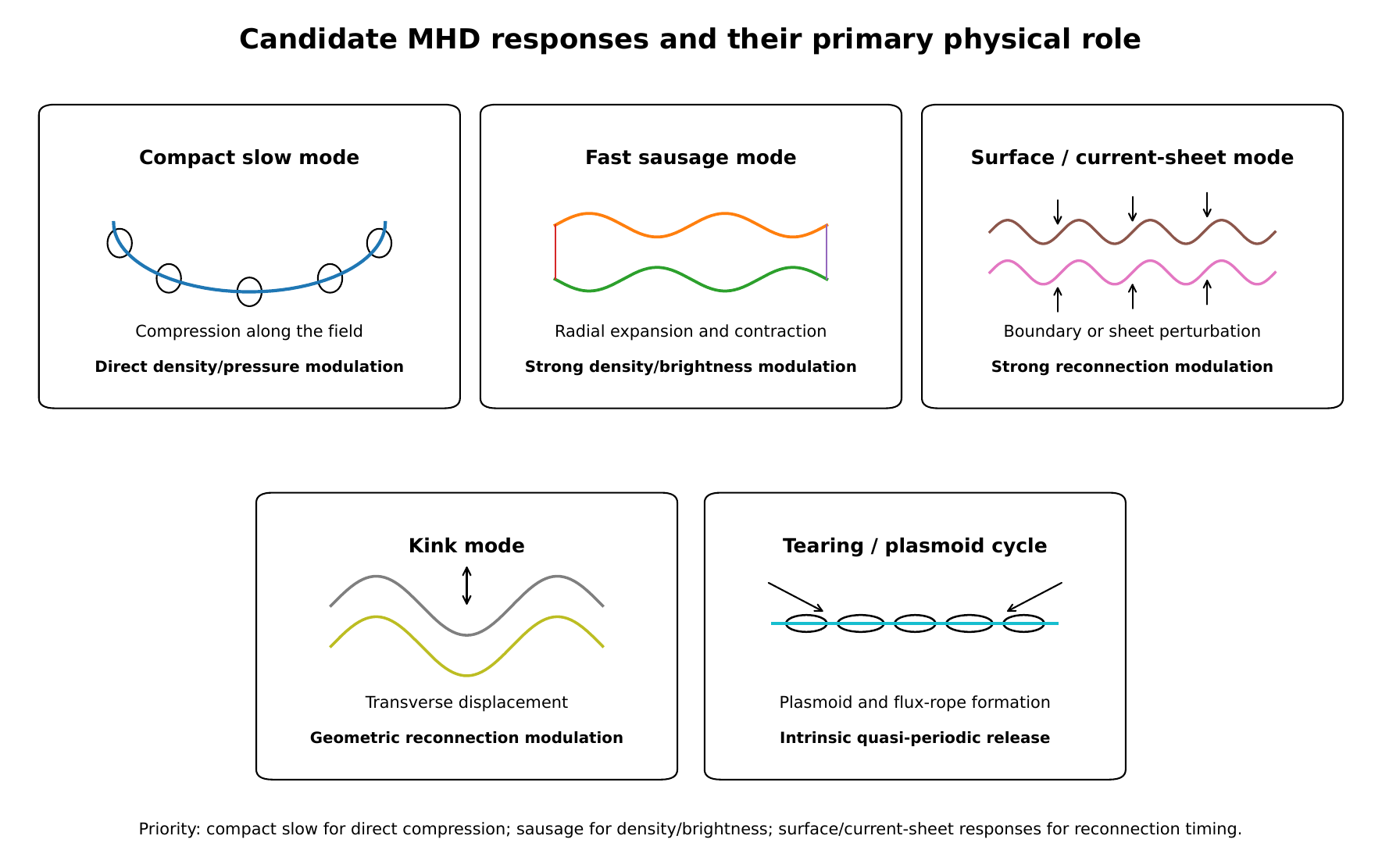}
\caption{Candidate MHD responses and their primary physical roles. Compact slow modes directly perturb density, pressure, and field-aligned velocity. Fast sausage modes perturb cross-sectional area, density, magnetic field, and white-light brightness. Surface and current-sheet modes act directly where reconnection occurs. Kink modes mainly displace the structure and can modulate reconnection geometrically. Tearing and plasmoid formation are nonlinear or unstable alternatives rather than stable eigenmodes.}
\label{fig:modes}
\end{figure*}

\subsection{Compact slow magnetoacoustic mode}

A slow mode is mainly field-aligned and compressive. It therefore provides a direct link to density and pressure modulation. The tube speed is
\begin{equation}
c_T=\frac{c_s v_A}{(c_s^2+v_A^2)^{1/2}},\qquad
c_s=\left(\frac{\gamma k_{\rm B}T}{\mu m_{\rm p}}\right)^{1/2}.
\end{equation}
Rather than fixing $c_T/c_s$ at one value, we use the sensitivity bracket $c_T/c_s=0.8$--$1.0$, which covers a moderate magnetic correction through the low-$\beta$ limit \citep{NakariakovVerwichte2005}. For $T=1$--$2$ MK, $L_{\rm eff}=0.2$--$1.0\Rsun$, and a field-aligned flow $U=0$--$50\ \kms$, the flow-corrected fundamental period from Equation~(\ref{eq:flow}) spans approximately $22$--$231$ min. Requiring $P=80$--$130$ min gives
\begin{equation}
L_{\rm eff}\simeq0.35\text{--}1.20\Rsun,
\label{eq:slowlength}
\end{equation}
where the bounds include the adopted temperature, tube-speed, and flow ranges. Thus the longer part of the observed interval can require a path slightly longer than the nominal $1\Rsun$ parameter interval. The result points to a sub-streamer cavity or an extended closed arcade below the cusp, rather than to a fundamental slow mode occupying the full visible length of a giant streamer.

A slow-mode interpretation must also address longitudinal reflection and damping. Thermal conduction can strongly damp standing slow modes in hot coronal structures \citep{OfmanWang2002}. We therefore use $\tau_{\rm d}/P\gtrsim2$ only as a practical survival criterion for modulation over several release events, not as a predicted slow-mode damping ratio. A quantitative value requires a specified density and temperature profile, reflecting boundaries, and a nonadiabatic eigenvalue calculation.

\subsection{Fast sausage mode}

The axisymmetric fast sausage mode changes the cross-sectional area of a magnetic waveguide. It can therefore produce strong perturbations of density, magnetic field, pressure, and line-of-sight brightness. This direct compressibility makes it an attractive candidate for coronagraphic density modulation.

The trapping condition is restrictive for a long, weakly contrasted streamer stalk. For a zero-$\beta$ top-hat cylinder with approximately equal internal and external magnetic field, the cutoff is \citep{EdwinRoberts1983}
\begin{equation}
k_{\rm c}a\simeq\frac{j_{0,1}}{\sqrt{\rho_i/\rho_e-1}},
\end{equation}
where $j_{0,1}=2.4048$. With $k_n\simeq n\pi/L$, the minimum trapped longitudinal harmonic is
\begin{equation}
n_{\min}=\left\lceil\frac{j_{0,1}}{\pi}\frac{L/a}{\sqrt{\rho_i/\rho_e-1}}\right\rceil.
\label{eq:nmin}
\end{equation}
For the adopted ranges $L/a=5$--$50$ and $\rho_i/\rho_e=1.6$--$2.5$, Equation~(\ref{eq:nmin}) gives $n_{\min}\simeq4$--$50$. Thus, the global fundamental sausage mode is not trapped anywhere in this deliberately slender, weak-contrast box. Trapped higher harmonics, shorter localized waveguide segments, smoother or current-carrying profiles, and leaky sausage responses remain possible \citep{Pascoe2007,Inglis2013,Vasheghani2014,LopinNagorny2019}.

For a long-wavelength leaky response, a useful dimensional transverse scale is
\begin{equation}
P_{\perp}\sim\frac{2\pi a}{j_{0,1}v_{\rm f}},
\label{eq:sausageperp}
\end{equation}
where $v_{\rm f}$ is a representative fast speed. We take an effective half-width $a=0.10$--$0.45\Rsun$, spanning narrow few-degree stalks and the approximately $600$ Mm full width reported for an individual COR1 streamer \citep{MorganCook2020,Srivastava2016}. For the speed benchmark we use $v_{\rm f}=250$--$530\ \kms$, which brackets the external Alfv\'en speeds inferred around a streamer plasma sheet between approximately $3$ and $5\Rsun$ after separating the solar-wind flow from the measured propagation speed \citep{Chen2011}. Equation~(\ref{eq:sausageperp}) then gives $P_{\perp}\simeq5.7$--$55$ min \citep{Cally1986,Vasheghani2014,LopinNagorny2019}. This is only a transverse fast-timescale benchmark. It is not a complete complex sausage eigenfrequency or leakage-time calculation, and $v_{\rm f}$ is not asserted to be the internal Alfv\'en speed of a single equilibrium.

\subsection{Surface and current-sheet modes}

A surface mode is localized near an interface rather than throughout the full volume. In a streamer environment, the relevant interface may be the edge of a dense plasma sheet, a separatrix, or a current sheet. These modes can be compressive or weakly compressive, depending on the branch, but their main advantage is location: they act where reconnection and plasma release occur.

For an initial scale comparison, we use the same outer-coronal fast-speed benchmark, $v_{\rm f}=250$--$530\ \kms$, and a local guided length of $0.1$--$1.0\Rsun$ \citep{Chen2011}. The resulting two-way scale is approximately $4.4$--$93$ min. This is explicitly not a current-sheet eigenmode solution and does not cover slower sheet branches or instability growth times. The actual period and damping depend on current-sheet thickness, magnetic and density contrasts, guide field, field-aligned wavenumber, background flow, velocity shear, and coupling to external continua. A full sheet dispersion relation is therefore required before a stable branch can be assigned a unique period or $\tau_{\rm d}/P$. Surface/current-sheet responses remain physically important because of their direct coupling to sheet thickness, reconnection inflow, and plasmoid release, not because the proxy period uniquely matches PDS.

\subsection{Kink mode as a geometric reconnection modulator}

The kink mode mainly displaces the axis of a tube, sheet, or streamer. In the linear limit it is only weakly compressive, so it is not the first candidate for direct density modulation. It can nevertheless change the position and curvature of the cusp, the angle between reconnecting fields, and the current-sheet geometry. Kink oscillations are therefore retained as geometric modulators of reconnection.

Streamer-wave observations establish that kink-like responses occur, but they also show why observer-frame propagation speeds should not be inserted directly into a standing-mode estimate. In a COR1 event at about $2\Rsun$, \citet{Srivastava2016} reported an event-specific transverse pulse with a period of approximately $25$ min, a propagation speed near $500\ \kms$, and a streamer full width of approximately $600$ Mm. In the outer corona, a survey of $22$ LASCO/COR2 streamer-wave events found periods of $2$--$8$ hr and measured outward phase speeds of $360$--$740\ \kms$ \citep{Decraemer2020Waves}. Those measured speeds contain a substantial flow contribution: comparison with $\lambda/P$ implied an average post-shock streamer-wind speed near $300\ \kms$. The same survey found that usually only one or two periods were visible.

We therefore constrain the kink candidate empirically rather than assigning it an artificial standing period or importing a coronal-loop resonant-absorption damping ratio. Figure~\ref{fig:survey} shows the approximately $25$ min COR1 event separately from the $120$--$480$ min global streamer-wave population. These observations demonstrate the existence of transient kink-like responses, but they do not establish a long-lived kink clock for PDS. Kink modes remain plausible geometric modulators of cusp or sheet geometry, while their streamer-specific leakage and damping require an open, flowing equilibrium.

\subsection{Torsional and Alfv\'enic nonlinear modulation}

A linear torsional Alfv\'en wave is nearly incompressible. It produces azimuthal velocity and magnetic-field perturbations but little first-order density modulation. Nonlinear magnetic-pressure effects, phase mixing, or coupling to compressive modes may nevertheless change local current formation and reconnection. We therefore treat torsional/Alfv\'enic disturbances as secondary nonlinear modulators rather than primary density modes.

\subsection{Tearing, plasmoid, and shear-driven alternatives}

Tearing and plasmoid formation are not stable resonant eigenmodes. They are intrinsic reconnection dynamics and can independently generate quasi-periodic release. Global MHD simulations of the heliospheric current sheet have produced short timescales of about $1$--$2$ hr and longer timescales of about $10$--$20$ hr through flow-modified tearing \citep{Reville2020}. Intrinsic streamer-instability models have also connected mixed sausage--kink behavior near the cusp to repeated reconnection and blob formation \citep{Chen2009}. These mechanisms provide an essential comparison for any externally or resonantly modulated model.

\section{Period Constraints and Parameter Space}

\subsection{Basic travel-time estimate}

For a uniform, stationary waveguide, the simplest standing-mode estimate is
\begin{equation}
P_n \simeq \frac{2L_{\rm eff}}{n v_{\rm ph}},
\label{eq:travel}
\end{equation}
where $n$ is the longitudinal harmonic number. Equation~(\ref{eq:travel}) is useful for an initial parameter map, but it should not be used as a mode identification by itself.

\subsection{Effects of stratification and background flow}

For a slowly varying structure with a field-aligned flow $U(s)$, a two-way travel-time estimate is
\begin{equation}
P_n \simeq \frac{1}{n}\left[
\int_0^L \frac{ds}{v_{\rm ph}(s)+U(s)}+
\int_0^L \frac{ds}{v_{\rm ph}(s)-U(s)}
\right].
\label{eq:flow}
\end{equation}
This expression requires $v_{\rm ph}>|U|$ along the return path. Near a critical point, the inward characteristic can become very slow or cease to provide a closed round trip. A sonic or Alfv\'enic critical region should therefore not be treated automatically as a rigid reflecting boundary.

\subsection{Parameters for the quantitative analysis}

Table~\ref{tab:parameters} lists the intervals used in the literature-based quantitative analysis. They do not define a single equilibrium model, and we therefore describe the analysis as observation-informed rather than observation-constrained. The $1$--$2$ MK sensitivity range brackets the nearly isothermal $1.43\pm0.08$ MK streamer plasma inferred above $1.2\Rsun$ by \citet{Goryaev2014}. The flow range $U=0$--$50\ \kms$ is consistent with the observational upper limit below approximately $2.5\Rsun$ \citep{Suess2006}. The inside/outside density contrast $1.6$--$2.5$ follows from paired COR1 density profiles between $2$ and $3\Rsun$ \citep{Srivastava2016}. The effective half-width is a geometric sensitivity bracket anchored by narrow few-degree stalks at $4$--$8\Rsun$ and an approximately $600$ Mm full-width COR1 event \citep{MorganCook2020,Srivastava2016}. The fast-speed benchmark $250$--$530\ \kms$ is anchored in flow-corrected outer-coronal streamer seismology \citep{Chen2011}. Global kink periods, observer-frame speeds, and train lengths are listed separately because the measured propagation speed includes solar-wind advection \citep{Decraemer2020Waves}.

\begin{table*}
\centering
\caption{Literature-anchored inputs and explicit sensitivity intervals. The entries do not define one equilibrium model.}
\begin{tabular}{llll}
\toprule
Structure & Quantity & Adopted value/range & Physical or observational basis \\
\midrule
Compact arcade/cusp cavity & $L_{\rm eff}$ & $0.2$--$1.0\Rsun$ & Geometric sensitivity \\
Compact arcade/cusp cavity & $T$ & $1$--$2$ MK & Brackets $1.43\pm0.08$ MK streamer plasma \\
Compact arcade/cusp cavity & $c_T/c_s$ & $0.8$--$1.0$ & Magnetic sensitivity bracket \\
Compact arcade/cusp cavity & $U$ & $0$--$50\ \kms$ & Upper limit below $\sim2.5\Rsun$ \\
Streamer stalk/plasma sheet & $a$ & $0.10$--$0.45\Rsun$ & Geometric sensitivity anchored by observed widths \\
Streamer stalk/plasma sheet & $L/a$ & $5$--$50$ & Slender-waveguide sensitivity \\
Streamer stalk/plasma sheet & $\rho_i/\rho_e$ & $1.6$--$2.5$ & Paired COR1 inside/outside densities \\
Fast-mode benchmark & $v_{\rm f}$ & $250$--$530\ \kms$ & Flow-corrected outer-coronal seismology \\
Surface/current-sheet segment & $L_s$ & $0.1$--$1.0\Rsun$ & Local-length sensitivity \\
Kink observations & $P$ & $\sim25$ min; $120$--$480$ min & One COR1 event; LASCO/COR2 survey \\
Kink observations & $v_{\rm obs}$ & $360$--$740\ \kms$ & Observer frame; includes flow \\
Kink observations & visible train & typically $1$--$2$ periods & Empirical lifetime constraint \\
\bottomrule
\end{tabular}
\label{tab:parameters}
\end{table*}

\subsection{Dispersion, cutoff, leakage, and damping}

A mode is dispersive when $\omega/k$ or $d\omega/dk$ depends on $k$. Different spectral components then propagate with different speeds and a wave packet changes shape \citep{Nakariakov2004}. A mode is trapped when the external perturbation decreases away from the waveguide. It is leaky when the external medium supports outgoing waves and energy escapes from the structure.

For a damped or leaky mode, the frequency is complex,
\begin{equation}
\omega=\omega_{\rm R}+i\omega_{\rm I},
\end{equation}
with
\begin{equation}
P=\frac{2\pi}{\omega_{\rm R}},\qquad
\tau_{\rm d}=\frac{1}{|\omega_{\rm I}|}.
\end{equation}
A mode with the correct period but $\tau_{\rm d}\ll P$ cannot organize a long train of plasma releases. The ratio $\tau_{\rm d}/P$ is therefore as important as the period itself.

\begin{figure*}
\centering
\includegraphics[width=0.98\textwidth]{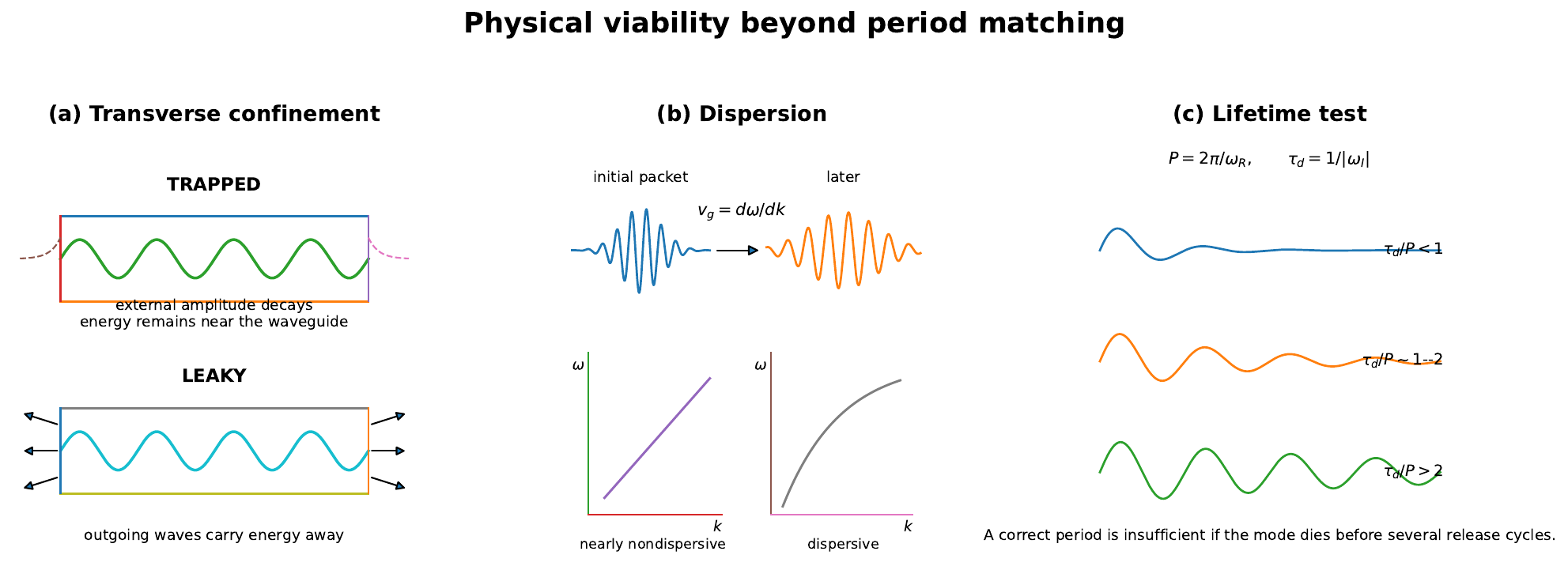}
\caption{Physical viability beyond period matching. (a) A trapped mode has an external perturbation that decays away from the waveguide, whereas a leaky mode radiates energy into the surrounding plasma. (b) In a dispersive waveguide, the group speed depends on wavenumber and an initially compact wave packet spreads and changes shape. (c) The damping ratio $\tau_{\rm d}/P$ controls whether a response disappears within one cycle, produces only a transient modulation, or survives for several possible release cycles.}
\label{fig:viability}
\end{figure*}

Figure~\ref{fig:viability} summarizes why period agreement alone is insufficient. Trapping tests the transverse confinement of wave energy, dispersion controls the evolution of a broadband disturbance, and the damping ratio tests whether a candidate can survive long enough to organize repeated plasma release.

\subsection{Quantitative assessment of candidate mechanisms}

\begin{figure*}
\centering
\includegraphics[width=0.98\textwidth]{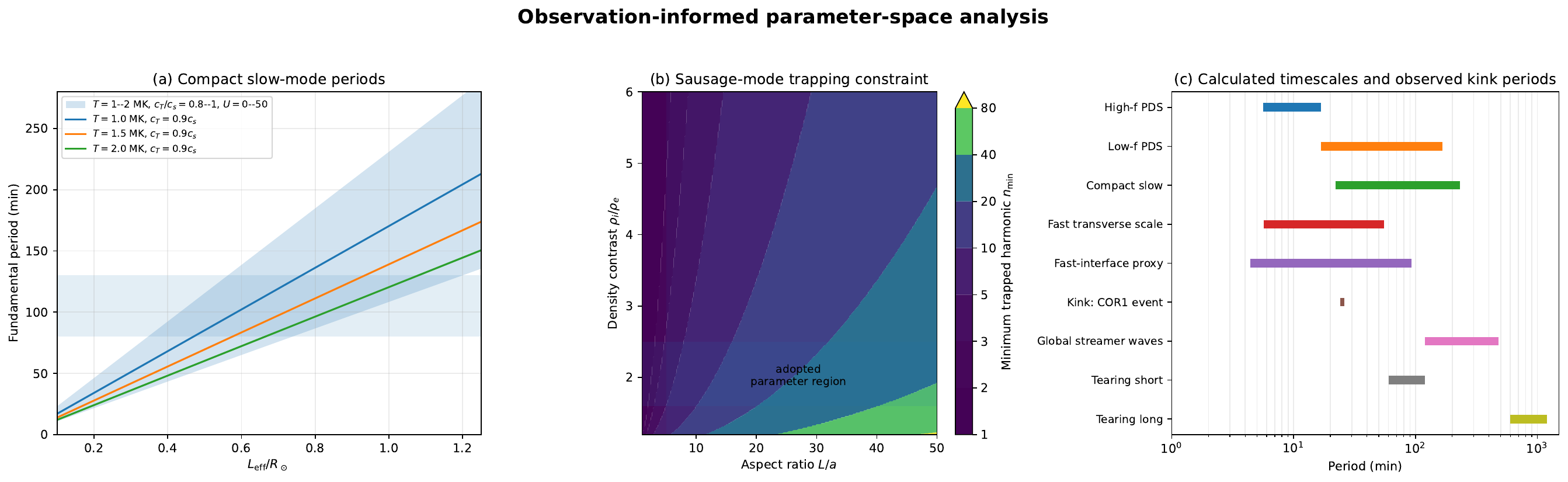}
\caption{Observation-informed parameter-space analysis. (a) Compact slow-mode period envelope for $T=1$--$2$ MK, $c_T/c_s=0.8$--$1.0$, and $U=0$--$50\ \kms$; representative $c_T=0.9c_s$ curves and the $80$--$130$ min band are also shown. (b) Minimum trapped sausage harmonic for a zero-$\beta$ top-hat cylinder; the shaded box covers $L/a=5$--$50$ and $\rho_i/\rho_e=1.6$--$2.5$. (c) Observed PDS populations, calculated sensitivity scales, and empirical kink constraints. The fast-interface bar is a proxy rather than a current-sheet eigenmode; the kink entries are observed periods rather than standing-mode calculations.}
\label{fig:survey}
\end{figure*}

Figure~\ref{fig:survey} replaces an intuitive period comparison with a quantitative parameter-space analysis. Compact slow modes can reach the low-frequency PDS population, but $80$--$130$ min requires $L_{\rm eff}=0.35$--$1.20\Rsun$ and sufficient reflection and lifetime. Fast sausage modes retain the strongest direct density signature, but the classical global fundamental is excluded by the trapping test in the adopted weak-contrast box; leaky or higher-harmonic responses remain possible. The fast-interface proxy overlaps both statistical PDS populations but does not establish a sheet eigenmode. The kink comparison is now empirical: a local COR1 pulse lies near $25$ min, whereas global LASCO/COR2 streamer waves occupy $2$--$8$ hr and typically persist for only one or two visible periods. Flow-modified tearing supplies an independent intrinsic alternative near $1$--$2$ hr and $10$--$20$ hr \citep{Reville2020}.

\begin{table}[t]
\centering
\caption{Quantitative assessment of the candidate mechanisms.}
\scriptsize
\setlength{\tabcolsep}{2pt}
\renewcommand{\arraystretch}{1.15}
\begin{tabular}{p{0.20\columnwidth}p{0.26\columnwidth}p{0.43\columnwidth}}
\toprule
Candidate & Quantitative constraint & Physical assessment \\
\midrule
Compact slow & $22$--$231$ min & Best-constrained stable compressive modulator among the candidates considered here; further assessment requires constraints on reflection and damping. \\
Surface/sheet & $4.4$--$93$ min fast proxy & High-priority coupling candidate; a sheet dispersion relation is required. \\
Fast sausage & $5.7$--$55$ min; $n_{\min}\simeq4$--$50$ & Conditional: leaky, local, or higher-harmonic response. \\
Kink & $\sim25$ min event; $120$--$480$ min survey & Observed but transient geometric modulator; usually $1$--$2$ visible cycles. \\
Torsional Alfv\'enic & No unique quantitative constraint & Secondary nonlinear candidate. \\
Tearing/\newline plasmoid & $1$--$2$ hr; $10$--$20$ hr & Strong intrinsic reconnection-driven alternative. \\
\bottomrule
\end{tabular}
\label{tab:ranking}
\end{table}

\section{Observable Discriminants}

A slow-mode interpretation predicts correlated density, pressure, temperature, and field-aligned velocity perturbations with mode-dependent phase relations. A sausage interpretation predicts changes in width and density, with corresponding modulation of white-light brightness. A kink interpretation predicts transverse displacement with weaker intrinsic compression, although line-of-sight effects may produce apparent intensity modulation. A surface or current-sheet response should be sought through changes in sheet position or thickness, cusp geometry, inflow, and the timing of plasmoid or blob release.

These tests require coordinated observations. Coronagraphic and heliospheric imaging can measure morphology and release timing. EUV and spectroscopic observations can constrain density, temperature, and velocity near the cusp. In situ measurements provide density, composition, magnetic field, and possible flux-rope signatures. No single observable is expected to identify the mode uniquely, but a combined phase and morphology test can reject incompatible interpretations. A particularly useful discriminator is whether a coherent source-region perturbation is phase-related to the timing of successive releases. Such phase coherence would support modulation, whereas a plasmoid sequence without a coherent upstream oscillatory signature would favor an intrinsic reconnection cycle.

\section{Discussion}

\subsection{Relation to S-Web and interchange-reconnection models}

The framework does not replace the standard reconnection-based picture of slow-wind release. It adds a possible temporal modulation layer. The proposed MHD responses do not compete with the S-Web or interchange-reconnection picture. Reconnection determines how plasma reaches open magnetic flux, whereas the present analysis asks whether the structured corona can introduce, select, or enhance preferred release timescales.

\subsection{Why consider an MHD modulator if reconnection can already be periodic?}

Reconnection and tearing can self-organize and produce plasmoid chains or repeated blob release. This intrinsic pathway is represented explicitly by the tearing/plasmoid candidate and does not require an external or resonant clock. The MHD-modulation pathway addresses a different possibility. In a structured streamer or current sheet, broadband driving or intermittent reconnection can excite responses whose periods and lifetimes depend on geometry, phase speed, trapping, leakage, and damping. Such responses may filter a broad driver, favor a limited range of timescales, or modulate the amplitude and location of reconnection.

The two pathways are complementary rather than competing: (1) intrinsically quasi-periodic reconnection and (2) reconnection whose cadence is modified by a structured MHD response. Observations may contain both, and different PDS populations may give different weights to the two pathways. The present analysis therefore does not claim that modulation is necessary. It tests when modulation is physically possible and observationally distinguishable. The decisive question is not whether reconnection and MHD responses coexist, but whether the latter leave a reproducible temporal or morphological imprint on the release process.

\subsection{Can several PDS populations have different origins?}

The two statistical PDS populations found by \citet{Kepko2024} have different composition behavior and may have different generation mechanisms. The high-frequency population may be connected to externally driven coronal oscillations. A direct comparison of the published occurrence distributions shows that the broad $1$--$3$ mHz PDS distribution, centered near $2.1$ mHz, is similar to the distribution of coronal transverse velocity fluctuations associated with p-mode conversion, which peaks near $2$ mHz. This distribution-level agreement makes p-mode-related coronal transverse fluctuations a promising candidate modulator for the high-frequency PDS population, but the causal pathway from coronal oscillations through reconnection to PDS still requires direct observational testing \citep{Morton2019,Kepko2024}. The lower-frequency population may be more closely connected with S-Web arcs, helmet streamers, and the heliospheric current sheet. Within each population, more than one MHD or reconnection process may contribute. The present framework is therefore intentionally plural rather than universal and includes both intrinsic reconnection cycles and MHD-modulated release.

\subsection{Analytical limitations and future MHD/forward modeling}

The analytical estimates presented here do not establish a unique mode identification. They provide a reproducible parameter-space analysis with explicit assumptions. The compact slow-mode result is a travel-time constraint, not a proof that suitable reflecting boundaries exist. The top-hat sausage test is deliberately restrictive and does not replace a complex-frequency solution for a smooth, expanding, current-carrying streamer. The transverse fast scale and the surface/current-sheet interval use a literature-anchored fast-speed benchmark but are not eigenmode solutions. For kink waves, observer-frame propagation speeds are not treated as intrinsic phase speeds because solar-wind advection is substantial. We therefore use the observed periods and short train lengths directly and do not import a coronal-loop damping formula into the open streamer geometry.

A definitive test will require realistic MHD simulations initialized with observationally constrained magnetic-field, density, temperature, flow, and current-sheet profiles. Such modeling is needed to determine excitation efficiency, transverse trapping, longitudinal reflection, mode coupling, leakage, damping, and nonlinear reconnection response. Synthetic EUV, white-light, and in situ observables are also required for direct comparison with measurements.

The absence of a full numerical model is not treated as evidence for or against a candidate. The purpose of the present analysis is to reduce an unconstrained numerical search to a focused set of testable configurations. In particular, future calculations should test a sub-streamer slow cavity, a leaky or higher-harmonic sausage response, and stable surface/current-sheet branches coupled to reconnection.

\section{Conclusions}

We propose an analytical source--modulator--output framework for quasi-periodic plasma release from coronal streamers. The main results are:
\begin{enumerate}
\item Reconnection associated with the S-Web, streamer cusp, or current sheet remains the plasma-release mechanism. The framework allows two complementary pathways: intrinsically quasi-periodic reconnection and reconnection whose timing or amplitude is modulated by an MHD response. Modulation is tested, not assumed.
\item For $T=1$--$2$ MK, $c_T/c_s=0.8$--$1.0$, and $U=0$--$50\ \kms$, compact slow-mode periods span about $22$--$231$ min. Reproducing $80$--$130$ min requires $L_{\rm eff}=0.35$--$1.20\Rsun$, pointing to a sub-streamer cavity or extended closed arcade rather than the full giant streamer.
\item In a classical top-hat streamer waveguide with $L/a=5$--$50$ and $\rho_i/\rho_e=1.6$--$2.5$, the fundamental fast sausage mode is not trapped. The minimum trapped harmonic is $n\simeq4$--$50$, while a literature-anchored transverse fast-timescale benchmark gives $5.7$--$55$ min.
\item Surface/current-sheet responses have the most direct stable-mode connection to reconnection timing, but their $4.4$--$93$ min interval is only a fast-interface proxy and a sheet-specific dispersion relation is required. Among the candidates considered here, compact slow modes are the best-constrained stable compressive modulators. Their further assessment requires constraints on longitudinal reflection and damping; sausage modes remain conditional because of the trapping constraint.
\item Kink-like disturbances are observationally established but should not be assigned a standing period from observer-frame propagation speeds. A COR1 event showed an approximately $25$ min pulse, while a LASCO/COR2 survey found global periods of $2$--$8$ hr and usually only one or two visible cycles. Their most plausible role here is transient geometric modulation rather than direct density modulation.
\item Tearing and plasmoid formation remain a high-priority intrinsic alternative, with simulated $1$--$2$ hr and $10$--$20$ hr timescales. An external or resonant modulator is therefore not required for every event. The present analysis supports a plural, testable framework rather than one universal MHD origin for all PDS.
\end{enumerate}
The next step is targeted, observation-constrained MHD and forward modeling of the physically motivated configurations.

\begin{acknowledgments}
The author is grateful to the Scientific Editor, Dr. Shadia R. Habbal, for her careful consideration of the manuscript, and to the anonymous referee for a thoughtful and constructive review that helped clarify the physical connections discussed here, including the possible role of p-mode-related coronal oscillations, and to state more clearly what is supported by the present analysis and what still requires further observational testing.

The author is grateful to members of the SOHO/LASCO and STEREO/SECCHI instrument teams, and especially to Angelos Vourlidas, for valuable discussions on streamer blobs, periodic density structures, and their coronagraphic signatures. The author also thanks Alexis Rouillard for discussions on periodic density structures and their heliospheric manifestations. Finally, the author is grateful to Valery M. Nakariakov for his careful and critical reading of earlier versions of the manuscript and for stimulating discussions on coronal MHD oscillations, waveguide physics, and coronal seismology.
\end{acknowledgments}
\bibliography{references}
\bibliographystyle{aasjournal}
\end{document}